# Coarse-grained Soft-Clusters Remain non-Diffusing in the Melt State

Qingzhi Zou[‡], Yihui Zhu[‡], Yifu Ruan[‡], Rui Zhang[⊥], GengXin Liu[‡,*]

[‡]Center for Advanced Low-dimension Materials, State Key Laboratory for Modification of Chemical Fibers and Polymer Materials, College of Material Science and Engineering, Donghua University, Shanghai 201620, China
[⊥]South China Advanced Institute for Soft Matter Science and Technology, School of Molecular Science and Engineering, South China University of Technology, Guangzhou, 510640, China

**Abstract**

Melts of 3-dimensional dendritic beads-springs, namely coarse-grained soft-clusters, are studied by molecular dynamics simulations. The goal is to elucidate the unique dynamics of giant molecules, or generally speaking, 3-dimensional architectured polymers. When constituted by more than the critical number around 200 beads, soft-clusters cannot diffuse or relax far above their glass transition temperature, although relaxation can happen on the level of beads. Each soft-cluster can only rotate in the cage formed by neighboring soft-clusters. Such a non-diffusing state would transform to the liquid state at exceptionally high temperature, e.g. 10 times the glass transition temperature. Agreeing with experiments, 3D hierarchies lead to unique dynamics, especially their divergent relaxation times with the number of beads. These unique dynamics are in sharp contrast with 1-dimensional chain-like polymers. We name such a special state as 'cooperative glass', because of the 'cooperation' of the 3D-connected beads. The design of soft-clusters may also resemble cooperative rearranging regions where cooperativeness is contributed by low temperature, thus offer further insights into the glass problem.

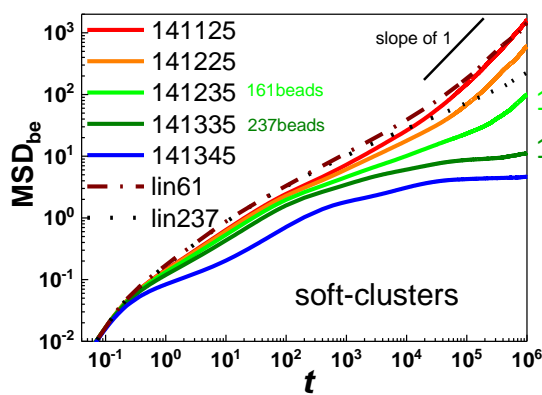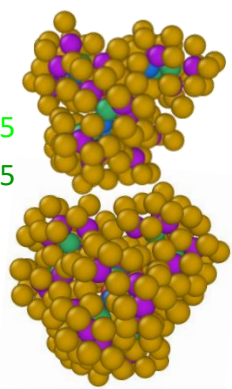

1. **Introduction**

Staudinger pointed out that polymers were chain-like one hundred years ago.[1] de Gennes identified the characteristic dynamics of high molecular weight chain-like polymers is reptation within a 1-dimensional (1D) tube formed by surrounding chains.[2-8] The terminal dynamics slows down to the 3.4 power of the molecular weight once they are above the entanglement molecular weight. However, a principle for the dynamics of 3-dimension architectured (3DA) polymers, with hierarchical internal architectures, remains missing. With certain degrees of 3D architectures, multi-arm star, polymer grafted nanoparticles, single-chain crosslinked nanoparticles (SCNP), and dendrimers, etc. have been well-studied.[9-11] However, multi-arm stars, for instance, have heterogeneous densities and can be modeled as a hard core and a soft corona.[12, 13] Their dynamics would be governed by the volume fraction of the hard core, and they may be modeled as one bead with specific interaction potentials. Thus, they may belong to soft colloids and are different from ideal 3DA polymers.

SCNPs would qualify as 3DA polymers, although have a disadvantage due to the intrinsical dispersity in their configurations. They have not been rheologically studied in the melt state because intra-polymer crosslinking at dilute concentrations cannot yield enough amount of samples. Thus, their dynamics are mostly studied as an additive in linear polymers. One molecular dynamics (MD) simulation on SCNP at the number density up to 0.7 reveal they can reach diffusion.[14] Due to synthetical challenges, dendrimers in experiments have limited numbers of generations. One exception is a 7$^{th}$ generation dendrimer ($M_w$ of 73 kg/mol) reported by Muzafarov et al.[15] It shows a liquid-to-solid transition and authors speculate it is due to the interpenetration.

Giant molecules were developed using molecular nanoparticles (such as POSS, polyhedral oligomeric silsesquioxane cages) as building blocks, and combing with precise synthesise.[16-20] Giant molecules can offer precise structures and compact 3D architecture, allowing the exploration of rich parameter space. Recent studies use giant molecules as model 3DA polymers and reveal a long-lasting elastic plateau with no relaxation when their diameter is larger than the critical diameter of 5 nm (or molecular weight $M_w$ of 38 kg/mol).[21-23] This leads to the hypothesis of cooperative glass, the non-diffusing state for melts of large enough 3DA polymers. As 3DA polymers grow larger and larger, e.g. from small molecules to infinite large 3D network, there must be a liquid-to-solid transition. The question would be where is the boundary and why there. Such a study would also contribute to establishing the boundary between colloidal domain and molecules/polymers, revealing when the dynamics would not be governed by the temperature. Existing theoretical understandings on colloids suspension and jamming are directly related to volume fraction, not diameters of 3DA polymer melts. Thus, further developments are needed to understand such solid-like dynamics and the sharp dynamics slow-down.

Linear polymers are modeled by linear chains of beads and springs. Here, we study beads-springs connected in 3D, namely the soft-cluster using MD simulation. The constituting beads would have their degree of freedom although constrained by springs. Regarded as a coarse-grained model of 3DA giant molecules, soft-clusters would reveal the consequence of this additional cooperativeness contributed by 3D connectivity and hierarchies, comparing to linear chains which have 1D connectivity and hierarchies. A clarification is needed to avoid confusion. Soft colloids are uncorrelated individual beads with variable interacting potentials.[24-26] Soft-clusters emphasize the synthesized correlation of multiple beads and belong to the molecular domain, not the colloidal domain.

2. **Simulation**

One type of construct can be labeled as $n_0n_1n_2n_3n_4n_5$ (141335, et al) and shown in Figure 1. Here, $n_0$ represents the center bead and is outward connected to $n_1$, $n_2$, $n_3$, $n_4$, and $n_5$ beads, respectively, by bond $b_{01}$, $b_{12}$, $b_{23}$, $b_{34}$, and $b_{45}$. The number of beads in the soft-cluster, $N_{be}$, is $1+n_1\times (1+n_2\times (1+n_3\times (1+n_4\times (1+n_5))))$. To prevent soft-clusters from degenerating to polymer chains or hairy stars, each $n_i$ should be large to maintain 3D architecture while not overcrowded. Such construct is in some way similar to dendrimers. However, previous simulations on dendrimers have not explored a large number (>150) of coarse-grained beads and high valency (>3) of beads [27-29].

The nonbonded interaction between beads in these soft-clusters is described by the truncated and shifted Lennard-Jones (LJ) potential,

$$U_{LJ}(r) = \begin{cases} 4\varepsilon\left[\left(\frac{\sigma}{r}\right)^{12} - \left(\frac{\sigma}{r}\right)^6 + \frac{1}{4}\right], & r \leq r_c = 2^{1/6}\sigma \\ 0, & r > r_c = 2^{1/6}\sigma \end{cases}$$

where $r$ is the distance between the center of two beads, $\sigma$ and $\varepsilon$ set the length and energy scale of the soft-cluster systems, respectively. $U_{LJ}(r)$, as shown in Figure S1(a), goes smoothly to zero at the cutoff distance $r_c$, which equals $2^{1/6}$ so that only repulsive interactions are included. The bond between two connected beads is the harmonic spring,

$$U_{Harm}(r) = \frac{k_b}{2}(r - r_0)^2 + U_{LJ}(r)$$

The equilibrium bond length $r_0$, equals 1.0. $k_b$ is the harmonic spring coefficient, and is set to 1000.

For simplicity, we set $\sigma$, $m$, $\varepsilon/k_B$ and $\sqrt{m\sigma^2/\varepsilon}$ as the units of length, mass, temperature, and time, respectively. Here, $m$ is the mass of beads and $k_B$ is Boltzmann constant. All variables and parameters are given in reduced units. The time step of $dt = 0.002$ is used to integrate Newtonian equations of motion. GPU-accelerated simulation package GALAMOST[30] is utilized to improve computational efficiency. The MD simulations are carried in a cubic box with periodic boundary conditions. Each simulation box contains 200 soft-clusters, so the total number of beads $N$ in the box is $200N_{be}$. To rule out any concern about the finite size of the simulation box, we verified on 141335 with 1000 soft-clusters, obtaining identical Mean Square Displacement (MSD). The MSD is a statistical measure of time-dependent particle displacements,

$$\Delta r^2(t) = \frac{1}{N}\langle \sum_i^N |r_i(t) - r_i(0)|^2 \rangle.$$

Soft-clusters are prepared at pressure $P=0.01$ and high temperature $T=10.00$. To ensure the equilibrated systems, we first perform MD simulations in the NPT ensemble for at least a production run of $10^4$ as illustrated in Figure S1(b). Then, we compress the systems into a target pressure ($P=10.00$) under constant temperature ($T=10.00$). After at least a production run of $10^4$, we subsequently cool the equilibrated systems from $T=10.00$ to $T=0.01$ at a rate of $10^{-4}$ and $P=10.00$. High pressure ensures the dense state. Under this cooling process, the number density, $\rho = N/V$, increases with decreasing temperature at constant pressure. Configurations obtained at desired temperatures are further equilibrated for at least 10 times longer than relaxation times of beads in the NVT ensemble before structural and dynamical properties are calculated. For each of $T$ we investigated, three independent simulations are performed to improve statistics.

The self part of intermediate scattering function $F_s$ is a statistical measure of time-dependent correlation functions to describe relaxation dynamics:

$$F_S(q,t) = \frac{1}{N}\langle \sum_i^N exp[i\boldsymbol{q} \cdot (\boldsymbol{r}_i(t) - \boldsymbol{r}_i(0))] \rangle,$$

where $N$ is the total number of beads, $q$ is the wave vector, and $r_i(t)$ is the position of $i$ bead at time $t$. It is the standard practice to determine the corresponding relaxation time as $F_s$ decays to 0.1.

Table 1. Properties of soft-clusters and one linear counterpart: $N_{be}$, number of beads in each soft-cluster; $T_g$, glass transition temperature; $T^{Co}$, cooperative transition temperature; at $T = 1$: $d_{sc}$, the diameter of soft-cluster; for linear counterparts, the table lists their $2R_g$;

$\tau_{be}$, the bead relaxation time; $\tau_{sc}$, the soft-cluster relaxation time; for linear counterparts; this is calculated at $q_{sc} \equiv \pi/R_g$.

| label | $N_{be}$ | $T_g^T$ | $T^{Co}$ | $d_{sc}$ | $\tau_{be}$ | $\tau_{sc}$ |
|---|---|---|---|---|---|---|
| 141125 | 61 | 0.43 | 0.56 | 4.4 | 3.8 | 1100 |
| 141225 | 113 | 0.44 | 0.60 | 5.3 | 4.0 | 3200 |
| 141235 | 161 | 0.44 | 0.74 | 6.0 | 5.3 | $2\times10^4$ |
| 141335 | 237 | 0.45 | 4.82 | 6.6 | 8.9 | $\gg 10^6$ |
| 141345 | 309 | 0.49 | 12.8 | 7.0 | 167 | $\gg 10^6$ |
| lin61 | 61 | 0.40 |  | 8.0 | 2.4 | 3900 |
| lin237 | 237 | 0.40 |  | 16.1 | 2.4 | $3\times10^5$ |

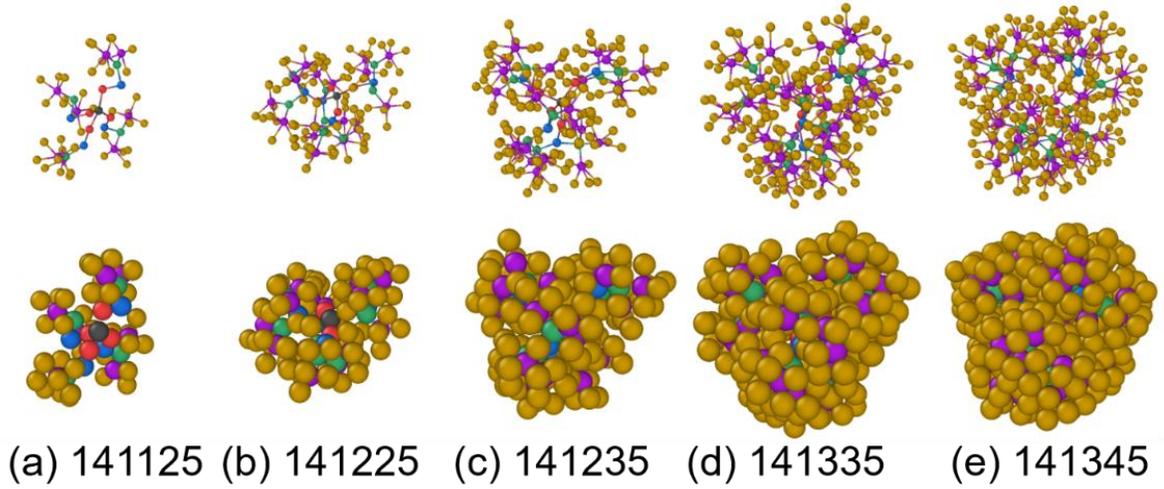

(a) 141125  (b) 141225  (c) 141235  (d) 141335  (e) 141345

Figure 1. Models illustrating each type of soft-clusters, $n_0n_1n_2n_3n_4n_5$. Six layers of beads from inside out are colored in gray, red, blue, green, purple, and brown.

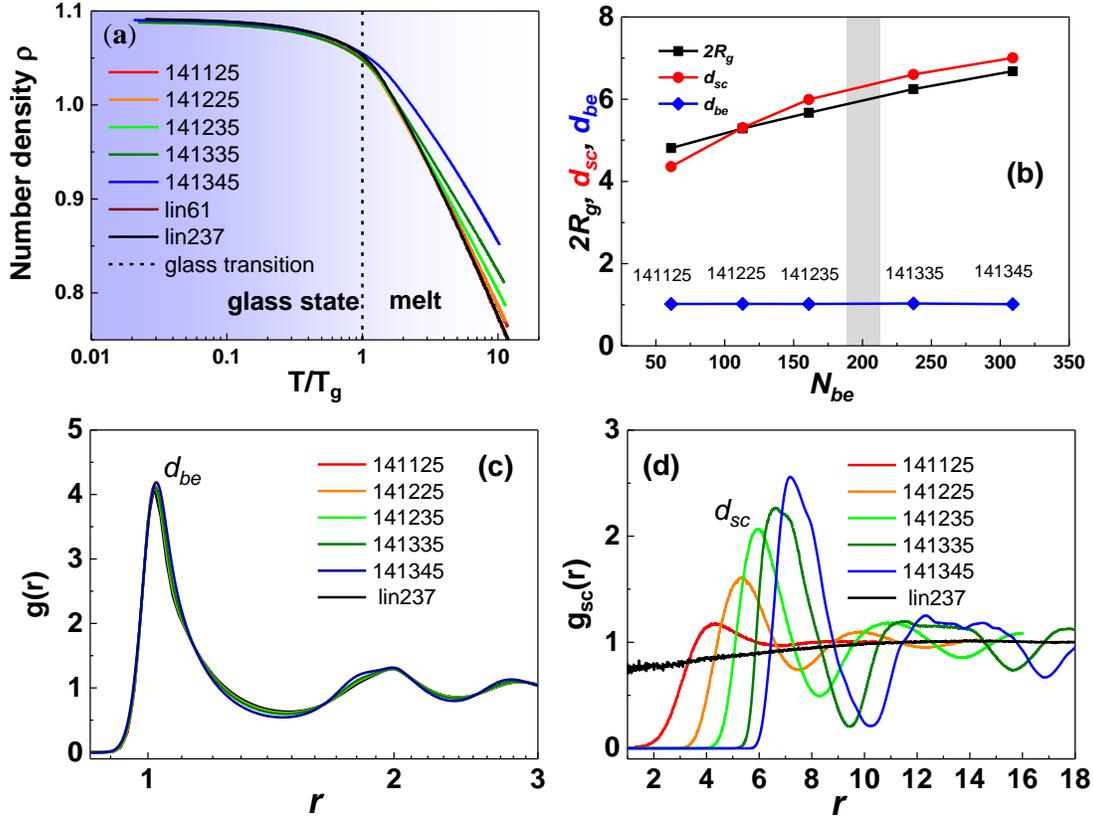

Figure 2. (a) The number density of various soft-clusters against normalized temperatures. At $T = 1$: (b) Diameters versus $N_{be}$; The radial distribution function of (c) all beads $g_{be}(r)$ and (d) the center of mass $g_{sc}(r)$. Their first peaks corresponding to $d_{be}$ and $d_{sc}$.

## 3. Results and Discussion

### 3.1 Glass transition and the static structure.

The conventional glass transition temperature $T_g$ may be thermodynamically identified from the slope change of specific volume $v_b \equiv 1/\rho$ vs. temperature, as shown in Figure S2(a) and listed in Table 1. The number density $\rho \equiv N/V$ is shown in Figure 2(a). Soft-clusters upto 141335 share similar $T_g$. In experimental studies, the upper-temperature limit is ~$1.5T_g$ before worrying about sample degradation [21-23]. Following simulations will first be performed at $T = 1$, which is about $2.2T_g$ and already higher than experimental temperature limit. Then we will explore the temperature dependencies of their dynamics.

Analysis of Voronoi neighbors in Figure S2(b) shows all soft-clusters have similar local environments. The distributions of bond lengths for each layer of bonds in each type of soft-cluster are given in Figures S2(c) and S3. There is negligible bond stretching up to 141335, whose inner bond length is 1.03 vs 1.02 of 141235. Figure 2(c) and (d) present the radial distribution function of all beads $g_{be}(r)$ and of the center of mass $g_{sc}(r)$ at $T = 1$. The current constructs are free from crystallization. The nearest neighbor peak in $g_{be}(r)$ equals 1, the bead diameter $d_{be}$. The diameter of the soft-cluster $d_{sc}$ is taken from the first peak in $g_{sc}(r)$. Diameters of soft-clusters are summarized in Figure 2(b), together with their radius of gyration $R_g$. As a reference, the linear counterpart lin61 and lin237 have the same number of beads as 141125 and 141335, and their $2R_g$ are 8.0 and 16.1, much larger than the diameters of soft-clusters.

### 3.2 The onset of 'cooperative glass' with increasing $N_{be}$.

In Figure 3(a), we present MSD$_{be}$, MSD averaged over all beads, at $T = 1$. 141235 and smaller soft-clusters achieve diffusion at $t > 10^5$, as indicated by the slope of 1. However, 141335 and larger soft-cluster show no sign of diffusion with MSDs remain as plateaus even at $t = 10^6$. In contrast, lin237 diffuses and relaxes faster than 141235 (which has 161 beads). The non-Gaussian parameter $a_2$ in Figure 3(b) is a description of the dynamic heterogeneity, whose magnitudes increase for larger soft-clusters. Non-zero $a_2$ proves the existence of 'cooperation', which is preset by the construct of soft-clusters.

F$_s$ as shown in Figures 3(c, d), 5, and S4 provide the corresponding relaxation times as F$_s$ decays to 0.1 (listed in Table 1). Two different wave vectors $q$ are evaluated: (1) F$_{s,be}$, Figure 3(c), using $q_{be}$ equals 7.2. $q_{be}$ comes from the first peak of the static structure factor (Figure S1(c)), and corresponds to $\tau_{be}$, the bead or segmental diffusion relaxation time. (2) F$_{s,sc}$, Figure 3(d), using $q_{sc} \equiv 2\pi/d_{sc}$ corresponds to $\tau_{sc}$, the soft-cluster or terminal diffusion relaxation time. $\tau_{sc}$ of 141335 becomes unmeasurable while its $\tau_{be}$ increases only 70% from 141235. $\tau_{sc}$ of linear counterparts are calculated at $q_{Rg} \equiv \pi/R_g$ as shown by dash-dot and dot lines and listed in Table 1. It is well known that the calculated relaxation times of linear polymers depends on the choice of $q$. If we calculate relaxation times of linear counterparts at $q_{sc} \equiv 2\pi/d_{sc}$, using $d_{sc}$ of 141125 and 141335, the results lead to 450 and 2200, much shorter than relaxation times at lower $q$.

The difference between 141335 and lin237 lies in the dimension in which beads are connected, 3D or 1D. This work verifies that large enough soft-clusters are indeed confined and reach a non-diffusing glass-like state. This glass-like state must be originated from cooperativeness, and we name it 'cooperative glass'. Next, we would analyze the temperature dependence.

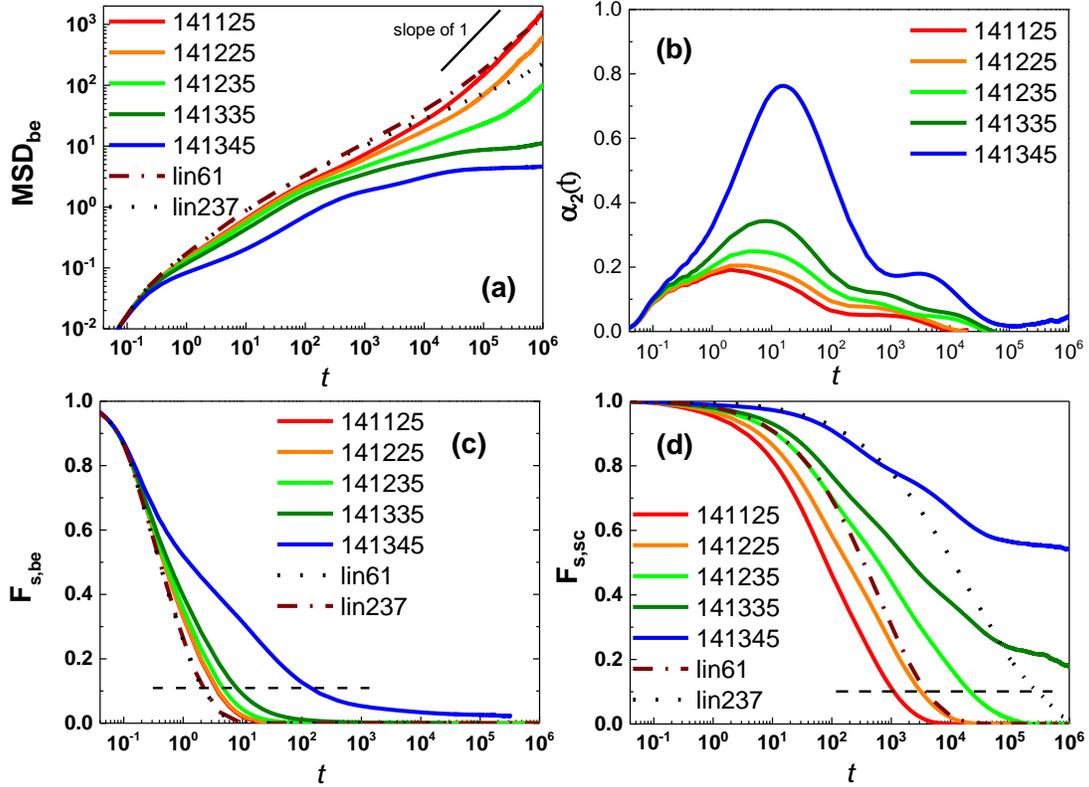

Figure 3. (a) MSD$_{be}$, MSD of all beads for different soft-clusters and lin237 at $T = 1$. (b) $a_2$, non-Gaussian parameters for different soft-clusters. (c) F$_{s,be}$, the self-part of the intermediate scattering function F$_s$ corresponding to $q_{be}$, and (d) F$_{s,sc}$, F$_s$ corresponding to $q_{sc}$. The relaxation time is taken from F$_s$=0.1 and yields $\tau_{be}$ and $\tau_{sc}$ as listed in Table 1.

### 3.3 Different regions in MSD.

MSD$_{be}$ and MSD$_{sc}$, MSD of the center of mass, at different temperatures for 141235 and 141335 are presented in Figure 4, while plots for all soft-clusters are given in Figure S6. The local minimums of the logarithmic derivative of MSD$_{be}$, Figure S7, correspond to cages at the respective length scale[31]. Three regions of dynamics can now be identified:

(1) BCA is the bead in the cage of neighboring beads. Beads start to move around and MSD$_{be}$ reaches 1. (2) CDI is the soft-cluster diffusion. At (although maybe unrealistically) high temperatures and long times, soft-clusters should reach free diffusion in this final region. MSD$_{be}$ and MSD$_{sc}$ both would be proportional to time, and their ratio equals 1. While 141235 or smaller soft-clusters can reach CDI, 141335 or even larger soft-clusters cannot reach CDI or can only reach CDI at unrealistic high temperature ($T > 10T_g$). (3) CRO as soft-cluster rotation in the cage of neighboring soft-clusters, and clearly seen in

141335 or even larger soft-clusters. This is at a higher temperature or longer time than BCA, but before reaching CDI. Even at long simulation times ($10^6$) and high temperatures ($T < 10T_g$), beads in large soft-clusters (141335 and beyond) cannot diffuse more than $d_{sc}$, but they would diffuse freely only within $d_{sc}$, so MSD$_{be}$ would plateau around $(d_{sc} - 1)^2$. The center of mass could not diffuse more than 1 ($d_{be}$). Thus, soft-cluster can only rotate, not diffuse, in this region.

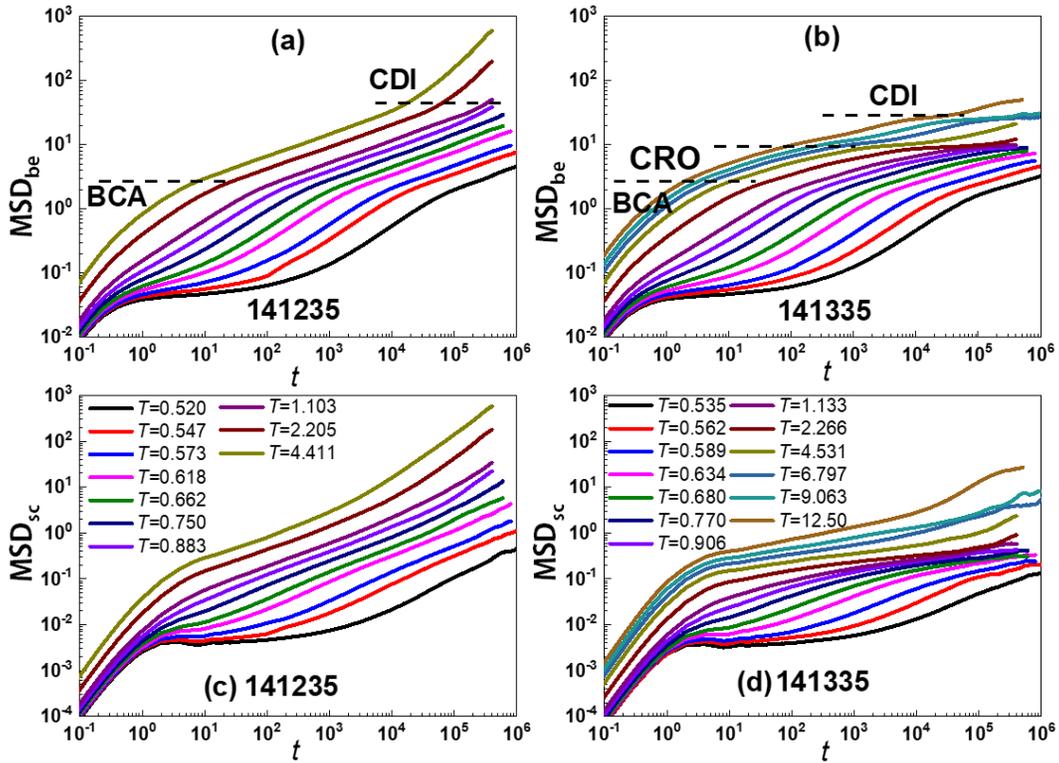

Figure 4 (a, b) MSD$_{be}$ and (c, d) MSD$_{sc}$ at different temperatures for (a,c) 141235 and (b,d) 141335. Temperatures are set from 1.18 to 10 times of $T_g$ for 141235, while going further up to 15, 20, and 27.6 times of $T_g$ for 141335. BCA is the bead in the cage of neighboring beads; CDI is the soft-cluster diffusion; CRO is soft-cluster rotation in the cage of neighboring soft-clusters.

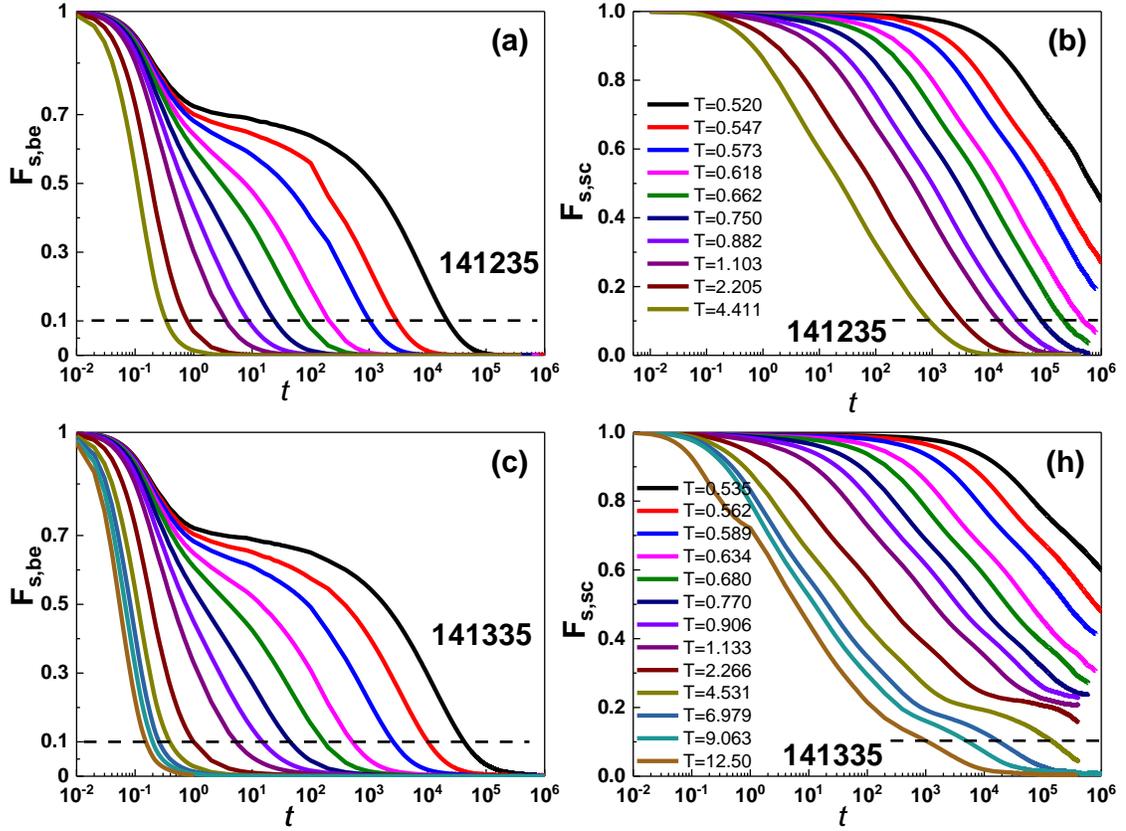

Figure 5. (a,c) $F_{s,be}$ and (b,d) $F_{s,sc}$ of soft-clusters (a,b) 141235 and (c,d) 141335 at various temperatures.

### 3.4 Relaxation time and the critical number $N_{cri}$.

The temperature dependency of $\tau_{be}$ and $\tau_{sc}$ are determined from $F_{s,be}$ and $F_{s,sc}$ (Figure 5 and S4) and summarized as symbols in Figure 6. Lines in Figure 6(a) and Figure S8 are Vogel-Fulcher-Tammann (VFT) fitting of $\tau_{be}$, revealing a universal good agreement except for extremely high temperatures. All five soft-clusters have similar $T_0$ around 0.39. From MSD we have already seen that 141335 could not reach diffusion, thus must have an extremely long terminal relaxation time. Indeed, $\tau_{sc}$ is unmeasurable at $T < 10T_g$. We use two points at the lowest temperatures with measurable $\tau_{sc}$ of 141335 to extrapolate to $T = 1$, and of 141345 to extrapolate to $T = 1$ and $10T_g$, suggesting $\tau_{sc}$ would be around $10^{14}$ for 141335 and $10^{36}$ for 141345 at $T = 1$, and $10^9$ for 141345 at $T = 10T_g$ (plotted by dotted lines in Figure 7).

Figure 7 presents relaxation times at two representative temperatures as a function of $N_{be}$. From 141235 to 141335, $N_{be}$ increases by 47%, $\tau_{be}$ increases by 70%. This is understandable because $\tau_{be}$ corresponds to rather localized dynamics. Even if the soft-cluster is not diffusing, relaxation at the bead level is not constrained. However, at a length scale about 7 times larger, $\tau_{sc}$ increases by probably $10^{10}$ times at $T = 1$, and by 500 times at $10T_g$ (unrealistically high temperature). $\tau_{sc}$ becomes decoupled with $\tau_{be}$. In this special state of cooperative glass, beads can diffuse and relax locally (liquid); yet at a larger length scale, soft-clusters cannot relax nor diffuse and are solid-like (glass).

This divergence of $\tau_{sc}$ is marked by a critical number of beads $N_{cri}$, around 200. Soft-clusters with more than $N_{cri}$ beads would only relax at unrealistic high temperatures in simulations. In experiments, molecular weight increases by 54% from OPOSS$_{16}$ to OPOSS$_{24}$ (OPOSS, octyl POSS), while the relaxation time increases by at least $10^8$ times [22, 23], similar to the present MD observation. In contrast, for linear polymers which slow down to 3.4 power of the molecular weight (the black line across lin237 in Figure 7), a ~50% increase of beads should only lead to a 4 times increase of relaxation time. The divergence of $\tau_{sc}$ can also be depicted by temperature. We define the cooperative transition temperature $T^{Co}$ by the temperature at which $\tau_{sc}$ reaches $10^5$ and plotted in Figure 8. It would separate cooperative glass and the liquid state, and also marks the onset of diffusion for soft-clusters. The experimental limit, as samples cannot endure arbitrary high temperature, of $1.5T_g$ is also provided and intersects at $N_{be}$ around 140. It suggests we would observe that relaxation times become unmeasurable for soft-clusters at $N_{cri}^{obs} = 140$.

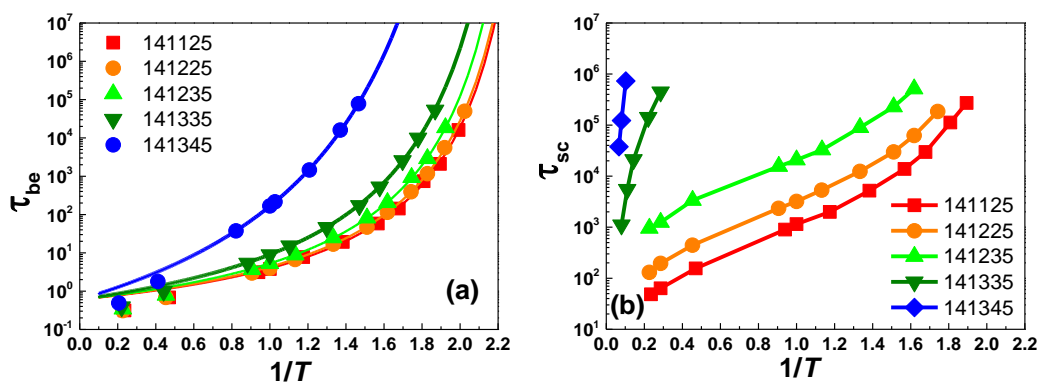

Figure 6. (a) The bead relaxation time $\tau_{be}$ and (b) the soft-cluster relaxation time $\tau_{sc}$ vs. inverse of temperature.

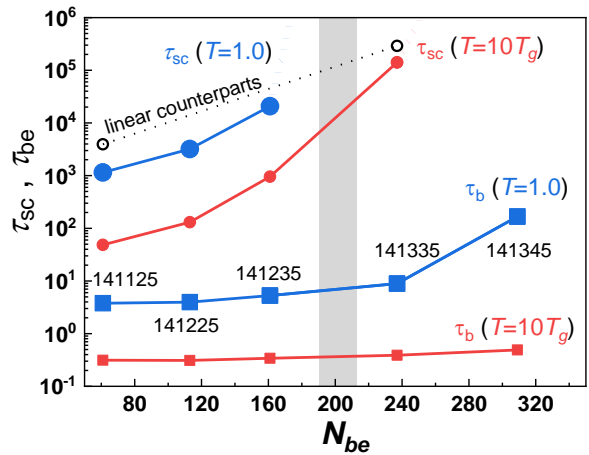

Figure 7. Relaxation times $\tau_{sc}$ (squares) and $\tau_{be}$ (circles) vs. $N_{be}$. Blue symbols are at $T = 1,$, which is about $2.2T_g$. From 141235 to 141335, $\tau_{be}$ increase by 70%, while $\tau_{sc}$, increase by about 10 orders of magnitude. A scaling around 3.4 for linear counterparts is provided as a reference.

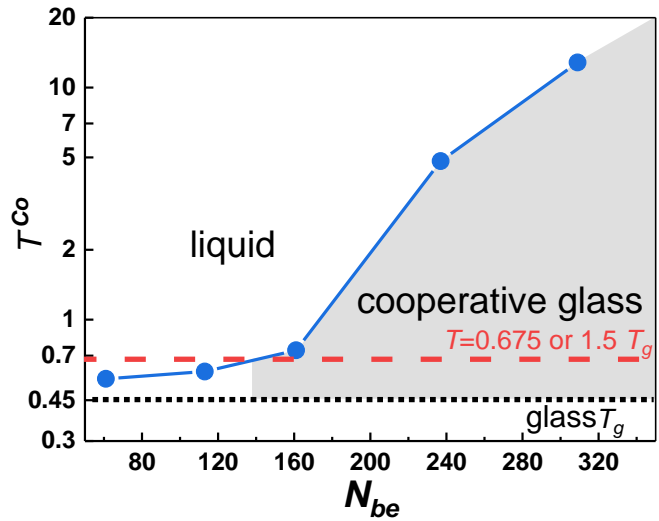

Figure 8. $T^{Co}$ vs. $N_{be}$; dashed line at $1.5T_g$ indicates the upper limit in experiments. Experimentally, soft-clusters with $N_{be}$ falls in the gray area would not relax before degradation happens.

**3.5 The cooperation.**

Analysis on OPOSS$_{16}$ and OPOSS$_{24}$ suggests that $N_{cri}^{obs}$ is around 110 [21, 22]. Simulations [31-34], experiments [35-38], and theoretical studies [39, 40] also hint at similar values. Such coincidence as well as the divergence of relaxation time with respect to $N_{be}$ would confirm the important role of 'cooperation' in 3DA beads-springs. The cooperative rearranging region (CRR) during glass transition refers to the concept that many beads would move cooperatively due to low temperature and lead to extremely slow dynamics.[41, 42] Soft-clusters with a fixed amount of cooperation may be analogous to CRR and offering further insight into the glass problem. Random First Order Transition theory deduced that the diameter of CRR, $\xi$, is proportional to $(\log\tau)^{2/3}$ around glass transition[40]. As $\xi \approx N_{be}^3$, then it suggests a scaling as indicated in Figure 4(c). It may overestimate the dynamic slow-down for small CRR and underestimate the dynamic slow-down for large CRR. We are inclined to use the critical number of beads to explain the onset of such cooperative glass transition. Alternatively, it may of merit to use elasticity or deformability to explain the observations.

4. **Conclusion**

"More is different."[43] said by P. Anderson. Now we find that in 3-dimension, more beads in soft-clusters, 3-dimensional architecture (3DA) beads-springs, indeed reveal different principles of dynamics. Specifically, "Giant is different."[19] Their dynamics dramatically slow down when soft-clusters contain more than the critical number of beads. Soft-cluster with 237 beads could not reach diffusion until at temperatures higher than $10T_g$. They revealed a special region of dynamics, soft-cluster rotation in the cage of neighboring soft-clusters, where beads can relax while soft-clusters cannot. We name this state 'cooperative glass'. It has characters of both liquid and glass and corresponds to the long-lasting elastic plateau in experiments on 3DA polymers. The study of 3DA polymers may also contribute to the glass problem and would establish the boundary between colloidal domain and molecules/polymers.

**Corresponding Author**
* Email:
GengXin Liu: lgx@dhu.edu.cn
**ORCID**
GengXin Liu (刘庚鑫): https://orcid.org/0000-0002-2998-8572

**Acknowledgments**
This work is supported by the Young Scientists Fund of the National Natural Science Foundation of China (Grant No. 21903013) and the Shanghai Municipal Education

Commission. The source codes and example dataset are available at: https://github.com/softcluster. Raw data are available upon request.

**Supplementary Materials**

Supplementary materials, Figures S1−S12 and Captions for supporting movies, associated with this article can be found, in the online version, at doi:10.1016/j.giant.2020.